\documentclass[12pt]{article}
\usepackage{graphicx} 
\usepackage{amsmath, amssymb, amsfonts, amsthm}
\usepackage{lipsum} 
\usepackage{setspace} 
\usepackage{geometry} 
\usepackage[export]{adjustbox}
\usepackage{hyperref} 

\title{Numerical Simulation of the Time-Dependent Schrödinger Equation Using the Crank-Nicolson Method}
\author{Adib Kabir}
\date{May 10, 2024}

\begin{document}

\maketitle
\begin{abstract}
This study presents a numerical simulation of a quantum electron confined in a 10 nm potential well, employing the Crank-Nicolson numerical technique to solve the time-dependent Schrödinger equation. Our results capture the evolution of the electron's wavefunction at the 2000th time step, illustrating distinct standing wave patterns and probability densities that validate quantum mechanical predictions. Additionally, both 2D and 3D simulations across multiple time steps reveal the dynamic nature of quantum superposition and interference within the well. These findings underscore the method's stability and accuracy, offering a robust tool for exploring quantum phenomena in constrained quantum systems
\end{abstract}

\section{Introduction}
Quantum mechanics revolutionizes our understanding of nature at the smallest scales. At its heart, the Schrödinger equation plays the most significant role, offering a framework to describe the dynamics of quantum systems. However, the complexity inherent in exact solutions, particularly for time-dependent equations, necessitates the development of robust numerical methods that can provide accurate predictions of quantum behaviors under various physical constraints.
\paragraph{} primary motivation for this project arises from the need to understand the time evolution of a quantum system with certain initial and boundary conditions. Traditional analytical methods often fall short of providing clear insights into such dynamics due to the non-linear and non-intuitive nature of quantum mechanics. As a result, there is a substantial need for better computational approaches that not only overcome these limitations but also enhance our ability to predict and manipulate quantum systems in technological applications such as quantum computing, cryptography, and tunneling phenomena.

\paragraph{}The main objective of this project is to predict the fate of an electron trapped inside a 10 nm box with the initial wavefunction equation of a Gaussian wave packet at time t = 0. In computational physics, solving a time-dependent Schrödinger equation numerically presents unique challenges, especially due to the need for stable and accurate solutions over time. The Crank-Nicolson method, a finite difference technique, is adept at addressing these challenges, offering a stable and second-order accurate solution in both space and time. To employ this method, we discretize the time and spatial domain in some discrete subintervals with uniform spacing. With the help of the Taylor series, we get the central finite difference equation to connect the initial condition that allows us to formulate a linear system connecting the initial wavefunction with the wavefunctions at any future time. After the formulation of the matrix, we simulate the real part, imaginary part, and probability density of the wavefunction over time. The conservation of probability and normalization conditions is investigated throughout the simulation, ensuring that the physical properties of the quantum system are accurately maintained.

\section{Physical Theory and Model}

\subsection{Time-Dependent Schrödinger Equation}
The time-dependent Schrödinger equation in one dimension for a free particle is given by:
\begin{equation}
    i\hbar \frac{\partial \Psi}{\partial t} = -\frac{\hbar^2}{2m} \frac{\partial^2 \Psi}{\partial x^2}
\end{equation}
where $\Psi(x,t)$ represents the wave function of the particle, $\hbar$ is the reduced Planck constant, and $m$ is the mass of the particle.

\subsection{Central Finite Difference Method}

Consider the function $\Psi(x)$ and we want to find the second derivative of this function at a point $x$. The second derivative is given by:

\begin{equation}
\frac{d^2\Psi}{dx^2} = \lim_{\Delta x \to 0} \frac{\Psi(x + \Delta x) - 2\Psi(x) + \Psi(x - \Delta x)}{(\Delta x)^2}
\end{equation}

Now, let's derive this discretization step by step:

\begin{enumerate}
    \item \textbf{Taylor Series Expansion}: Expand $\Psi(x + \Delta x)$ and $\Psi(x - \Delta x)$ in a Taylor series around the point $x$:
    
    \begin{equation}
    \Psi(x + \Delta x) = \Psi(x) + \Delta x \frac{d\Psi}{dx}\Bigg|_{x} + \frac{(\Delta x)^2}{2} \frac{d^2\Psi}{dx^2}\Bigg|_{x} + \mathcal{O}((\Delta x)^3)
    \end{equation}
    
    \begin{equation}
    \Psi(x - \Delta x) = \Psi(x) - \Delta x \frac{d\Psi}{dx}\Bigg|_{x} + \frac{(\Delta x)^2}{2} \frac{d^2\Psi}{dx^2}\Bigg|_{x} - \mathcal{O}((\Delta x)^3)
    \end{equation}
    
    Here, $\mathcal{O}((\Delta x)^3)$ denotes the remainder terms in the Taylor series expansion, which become negligible as $\Delta x$ becomes very small.
    
    \item \textbf{Subtract the Equations}: Subtract $\Psi(x - \Delta x)$ from $\Psi(x + \Delta x)$ to eliminate the first derivative term:
    
    \begin{equation}
    \Psi(x + \Delta x) - \Psi(x - \Delta x) = 2\Delta x \frac{d\Psi}{dx}\Bigg|_{x} + \mathcal{O}((\Delta x)^3)
    \end{equation}
    
    Since we are subtracting, the first derivative terms cancel out, and the odd-powered remainder terms also cancel out, leaving us with only the even-powered terms, specifically the second derivative term.
    
    \item \textbf{Construct the Second Derivative Approximation}: Add $\Psi(x - \Delta x)$ and $-2\Psi(x)$ to both sides and rearrange the terms:
    
    \begin{equation}
    \Psi(x + \Delta x) - 2\Psi(x) + \Psi(x - \Delta x) = (\Delta x)^2 \frac{d^2\Psi}{dx^2}\Bigg|_{x} + \mathcal{O}((\Delta x)^4)
    \end{equation}
    
    \item \textbf{Final Discretized Equation}: Now, we neglect the higher-order terms $\mathcal{O}((\Delta x)^4)$ assuming $\Delta x$ is small, and we get the central difference approximation for the second derivative:
    
    \begin{equation}
    \frac{d^2\Psi}{dx^2}\Bigg|_{x} \approx \frac{\Psi(x + \Delta x) - 2\Psi(x) + \Psi(x - \Delta x)}{(\Delta x)^2}
    \end{equation}
    
\end{enumerate}

So we have approximated the continuous second derivative with discrete points, which can be computed numerically. This is the basis for the finite difference method used to numerically solve the time-independent second derivative part of the Schrödinger equation.

\subsection{Crank-Nicolson Discretization for the Schrödinger Equation}

The one-dimensional, time-dependent Schrödinger equation for a particle with potential energy $V(x)$ is given by:
\begin{equation}
    i\hbar \frac{\partial \Psi}{\partial t} = -\frac{\hbar^2}{2m} \frac{\partial^2 \Psi}{\partial x^2} + V(x) \Psi
\end{equation}

Now, let's discretize the problem and derive the Crank-Nicolson method:

We discretize the spatial domain into points $x_j = j\Delta x$ for $j=0,1,\ldots,N$ and the temporal domain into points $t_n = n\Delta t$ for $n=0,1,\ldots,M$. The wave function at these discrete points is represented as $\Psi_j^n$.

For the time derivative at the midpoint between $t_n$ and $t_{n+1}$, we use the following finite difference approximation:
\begin{equation}
    \frac{\partial \Psi}{\partial t}\Bigg|_{t_{n+\frac{1}{2}}} \approx \frac{\Psi_j^{n+1} - \Psi_j^n}{\Delta t}
\end{equation}

For the spatial second derivative, we use the central difference at times $t_n$ and $t_{n+1}$, averaging them to center the spatial derivative in time:
\begin{equation}
    \frac{\partial^2 \Psi}{\partial x^2}\Bigg|_{t_{n+\frac{1}{2}}} \approx \frac{1}{2}\left( \frac{\Psi_{j+1}^{n+1} - 2\Psi_j^{n+1} + \Psi_{j-1}^{n+1}}{\Delta x^2} + \frac{\Psi_{j+1}^n - 2\Psi_j^n + \Psi_{j-1}^n}{\Delta x^2} \right)
\end{equation}

We substitute these finite difference expressions into the Schrödinger equation, yielding a discretized form that connects two time levels:
\begin{equation}
    i\hbar \frac{\Psi_j^{n+1} - \Psi_j^n}{\Delta t} = -\frac{\hbar^2}{4m}\left( \frac{\Psi_{j+1}^{n+1} - 2\Psi_j^{n+1} + \Psi_{j-1}^{n+1}}{\Delta x^2} + \frac{\Psi_{j+1}^n - 2\Psi_j^n + \Psi_{j-1}^n}{\Delta x^2} \right) + \frac{V_j}{2} (\Psi_j^{n+1} + \Psi_j^n)
\end{equation}

To facilitate solving this equation, we represent it in matrix form. Define matrices $T_1$ and $T_2$ based on the spatial discretization of the kinetic and potential energy terms:
\begin{align}
    T_1 &= \frac{\hbar^2}{2m\Delta x^2} \begin{pmatrix}
        2 & -1 & 0 & \cdots & 0 \\
        -1 & 2 & -1 & \cdots & 0 \\
        0 & -1 & 2 & \ddots & \vdots \\
        \vdots & \vdots & \ddots & \ddots & -1 \\
        0 & 0 & \cdots & -1 & 2
    \end{pmatrix} \\
    T_2 &= \text{diag}(V_1, V_2, \ldots, V_N)
\end{align}

The Crank-Nicolson method then takes the form:
\begin{equation}
    (I + \frac{i\hbar \Delta t}{2\hbar^2}T_1)\Psi^{n+1} = (I - \frac{i\hbar \Delta t}{2\hbar^2}T_1)\Psi^n + \Delta t T_2 \Psi^n
\end{equation}
where $I$ is the identity matrix of the same dimension as $A$.

Our problem involves a particle trapped in a potential well of length $L$ where $V(x)= \infty$ at the boundaries of the finite well, and zero at rest other positions in between. At time $t=0$, the initial wave function of the particle looks like:
\begin{equation}
\Psi(x,0) = Ae^{-\frac{(x-x_0)^2}{4\sigma^2}} e^{ikx}
\end{equation}
where:
\begin{itemize}
    \item $A$ is a normalization constant,
    \item $x_0$ is the initial position of the center of the wave packet,
    \item $\sigma$ specifies the width of the wave packet,
    \item $k$ is the initial wave number, related to the particle's momentum.
    \item $M$ is the mass of the particle
\end{itemize}
The above equation (11) can now be expressed in the following form: 
\begin{equation}
    i\hbar \frac{\Psi_j^{n+1} - \Psi_j^n}{\Delta t} = -\frac{\hbar^2}{4m}\left( \frac{\Psi_{j+1}^{n+1} - 2\Psi_j^{n+1} + \Psi_{j-1}^{n+1}}{\Delta x^2} + \frac{\Psi_{j+1}^n - 2\Psi_j^n + \Psi_{j-1}^n}{\Delta x^2} \right) 
\end{equation}

In this case, we must get the following matrix form, where the matrices $A$ and $B$ are defined for the Crank-Nicolson method:
\begin{align}
    A &= I + \frac{i\hbar \Delta t}{2m\Delta x^2} T \\
    B &= I - \frac{i\hbar \Delta t}{2m\Delta x^2} T
\end{align}
where $T$ is the tridiagonal matrix:
\begin{equation}
    T = \frac{\hbar^2}{2m\Delta x^2} \begin{pmatrix}
        2 & -1 & 0 & \cdots & 0 \\
        -1 & 2 & -1 & \cdots & 0 \\
        0 & -1 & 2 & \ddots & \vdots \\
        \vdots & \vdots & \ddots & \ddots & -1 \\
        0 & 0 & \cdots & -1 & 2
    \end{pmatrix}
\end{equation}
and $I$ is the identity matrix of the same dimension as $T$.

Boundary conditions are applied as:
\begin{equation}
    \Psi_0^n = \Psi_N^n = 0 \quad \text{for all } n
\end{equation}

The equation system then solves for $\Psi^{n+1}$ from:
\begin{equation}
    A\Psi^{n+1} = B\Psi^n
\end{equation}

The matrix equation (represents a linear system that can be efficiently solved using tridiagonal matrix algorithms, given the initial condition $\Psi^0$ and applying the boundary conditions at each time step.

\section{Algorithm for solving time-dependent Schrödinger equation}

The simulation program is designed to demonstrate the time evolution of the wavefunction of an electron in a one-dimensional potential well using the Crank-Nicolson method, a numerical scheme that ensures stability and accuracy by being unconditionally stable and second-order accurate in both time and space.

\subsection{Constants and Parameters:}
\begin{itemize}
    \item $L = 1 \times 10^{-8}$ meters (Length of the potential well).
    \item $N = 1000$ (Number of spatial divisions within the well).
    \item $a = L / N$ (Spatial grid size).
    \item $\hbar = 1.0545718 \times 10^{-34}$ Joule-seconds (Reduced Planck's constant).
    \item $m = 9.10938356 \times 10^{-31}$ kilograms (Mass of an electron).
    \item $h = 1 \times 10^{-18}$ seconds (Time step for the simulation).
\end{itemize}

\subsection{Initial Setup:}

\begin{itemize}
    \item \textbf{Wave Function Initialization,} $\psi_0$: The wavefunction is initialized using a Gaussian wave packet centered at $x_0 = L/2$ with a width $\sigma = 1 \times 10^{-10}$ meters and a wave number $k = 5 \times 10^{10}$. This initial wavefunction is normalized to ensure its integral over the spatial domain equals one.
\end{itemize}

\subsection{Matrix Setup with Crank-Nicolson Co-efficient:}

\begin{itemize}
\item $a_1$: Coefficient for the diagonal elements in matrix $A$, calculated as $1 + i \frac{\hbar h}{2ma^2}$.

    \item $a_2$: Coefficient for the off-diagonal elements in matrix $A$, calculated as $-i \frac{\hbar h}{4ma^2}$.
    \item $b_1$: Coefficient for the diagonal elements in matrix $B$, similar to $a_1$ but with a negative sign in the imaginary part, calculated as $1 - i \frac{\hbar h}{2ma^2}$.
    \item $b_2$: Coefficient for the off-diagonal elements in matrix $B$, opposite sign to $a_2$, calculated as $i \frac{\hbar h}{4ma^2}$.
\end{itemize}

\subsection{Matrix Construction}

\subsubsection{Upper and Lower Diagonals}
\begin{itemize}
    \item \textbf{A\_upper} and \textbf{A\_lower} are arrays filled with the value of $a_2$. These represent the off-diagonal entries in the matrix $A$, which handle the second spatial derivatives in the Crank-Nicolson scheme. The off-diagonal terms introduce the coupling between neighboring spatial points in the wavefunction.
    \item \textbf{B\_upper} and \textbf{B\_lower} are similarly set up using the value of $b_2$. The sign difference between $a_2$ and $b_2$ accounts for the forward and backward time-stepping in the implicit method.
\end{itemize}

\subsubsection{Main Diagonal}
\begin{itemize}
    \item \textbf{A\_diag} is an array filled with the value of $a_1$, placed along the main diagonal of matrix $A$. This term includes the effect of the potential energy at each point and balances the kinetic contributions from the off-diagonal terms.
    \item \textbf{B\_diag} uses $b_1$ and serves a similar purpose in matrix $B$ but corresponds to backward stepping in time, thereby completing the second half of the Crank-Nicolson temporal coupling.
\end{itemize}

\subsubsection{Banded Matrix Representation}
\begin{itemize}
    \item \textbf{A\_banded} and \textbf{B\_banded} are constructed to fit the required input format for the \texttt{solve\_banded} function from \texttt{scipy.linalg}. This format optimizes the solving of tridiagonal systems by focusing only on the non-zero diagonals. For matrix $A$:
    \begin{itemize}
        \item The top row consists of zeros followed by all $A\_upper$ elements.
        \item The middle row starts with $A\_diag$ elements.
        \item The bottom row ends with zeros following all $A\_lower$ elements.
    \end{itemize}
    \item Matrix $B$ is constructed similarly but uses $B\_upper$, $B\_diag$, and $B\_lower$.
\end{itemize}

\subsection{Time Evolution Computation}

\begin{itemize}
    \item \texttt{solve\_banded()}: Function from \texttt{scipy.linalg} used to solve the system of equations derived from the Crank-Nicolson method.
    \item Boundary conditions: Applied at the edges of the well (psi values at the first and last positions set to zero) to simulate infinite potential barriers.
    \item Time evolution loop: Executes over 2000 steps, updating the wave function at each time step using the Crank-Nicolson formula.
\end{itemize}

\subsection{Static Visualization}

\begin{itemize}
    \item \textbf{2D Visualization}: The real, imaginary, and probability density parts of the wavefunction are plotted at selected time steps such as 500th, 1000th, 1500th, and 2000th to illustrate the wavefunction's future evolution.
    \item \textbf{3D Visualization}: The 3D plot shows the complex nature of the wavefunction, combining real and imaginary components across the selected time steps.
\end{itemize}

\subsection{Animations}

\textbf{a) 2D Animation:} Separate animations for the real part, imaginary part, and probability density of the wavefunction are created using \texttt{FuncAnimation}. Each frame of the animation updates the wavefunction's profile to reflect its state at a new time step:
    \begin{itemize}
        \item Each frame shows the wavefunction's respective component along the spatial axis.
        \item \texttt{xlim} and \texttt{ylim} are set to appropriate ranges to maintain consistent visual scales.
        \item The animation progresses through each timestep, illustrating how the wavefunction evolves from its initial state to its state at the end of the simulation.
    \end{itemize}

\textbf{b) 3D Animation:} The 3D animations combine the real and imaginary parts of the wavefunction to provide a comprehensive view of its complex nature over time.
    \begin{itemize}
        \item The 3D plot setup uses \texttt{Axes3D} for three-dimensional plotting.
        \item Real and imaginary components of the wavefunction are plotted against the spatial dimension, providing a dynamic representation of the wavefunction's evolution.
        \item The \texttt{animate\_3D} function updates both the real and imaginary components in sync with each frame, allowing the viewer to observe the intertwined dynamics of these components.
    \end{itemize}

\subsection{Libraries and Modules}

The simulation utilizes several Python libraries and modules, each serving distinct roles in numerical calculations, data visualization, and dynamic animation creation. Here we detail the purpose of each module within the simulation process.

\paragraph{}
\textbf{a) NumPy:} This library is employed for its powerful numerical array structures and operations. It facilitates the efficient handling of large arrays and matrices, which are fundamental in performing the complex computations required for solving the quantum mechanical equations.
\paragraph{}
\textbf{b) MatPlotLib:} It is used for creating static, interactive, and animated visualizations in Python. This library allows for the plotting of the wavefunction's real, imaginary, and probability density aspects over time. It provides an intuitive interface for generating plots and figures, essential for analyzing and presenting the simulation results.
\paragraph{}
\textbf{c) SciPy:} This python library, specifically its \texttt{linalg} module, provides the \texttt{solve\_banded} function crucial for efficiently solving the banded systems of linear equations generated by the Crank-Nicolson method. This capability is vital for the iterative computation of the wavefunction at each time step.
\paragraph{}
\textbf{d) MPL\_toolkits.mplot3d:} MPL\_toolkits.mplot3d, a toolkit extension of Matplotlib, is used for generating three-dimensional plots. This module is critical for visualizing the complex structure of the wavefunction in a three-dimensional space, enhancing the interpretability of its dynamical changes over time.
\paragraph{}
\textbf{e) FuncAnimation:} FuncAnimation from Matplotlib's \texttt{animation} module plays a pivotal role in creating animations that depict the evolution of the wavefunction over time. These animations provide a dynamic visual representation of the simulation, illustrating the quantum behavior of the particle under study.

\section{Simulation Results}
In this section, we discuss our simulation results for the final time step and multiple time steps. 
\paragraph{}
To begin with, the real part of the wavefunction at the 2000th time step in left diagram of Figure 1 showcases a pattern of standing waves characterized by nodes and antinodes. This oscillatory behavior reflects the quantum mechanical principle that the wavefunction can exhibit both constructive and destructive interference patterns within a confined space. The peaks and troughs represent the amplitude of the wavefunction where the probability of finding the electron varies along the well. Similarly, the imaginary part exhibits oscillatory characteristics, which is expected as it represents another aspect of the wavefunction's complex nature. The presence of both real and imaginary components underscores the wave-like nature of quantum mechanical entities, which are described by solutions to the Schrödinger equation.

\paragraph{}
The probability density, calculated as the square of the absolute value of the wavefunction, provides the most direct physical insight into the electron's behavior within the well. The probability density graph in Figure 1's left diagram shows a large amplitude nearly around the center of the well, tapering off towards the edges. Such a distribution suggests that the electron is most likely to be found towards the center of the potential well, which is consistent with the quantum mechanical behavior of particles in a "box" or potential well where boundary conditions force the wavefunction to zero at the edges.

\paragraph{}
The 3D plot of the wavefunction at the final step combines the real and imaginary parts to form a spatial representation of the wavefunction's complex amplitude in three dimensions. This visualization helps in understanding the multidimensional aspect of quantum states, which is not apparent from 1D or 2D plots. The graph shown in the right diagram of Figure 1 reflects the combined oscillations of the real and imaginary components, highlighting the central region as the region highest probability of locating the electron. At 2000th step, the electron is spreading like a wave throughout the whole potential well. This suggests that at this instant, the electron has less particle property but more wave property.

\begin{figure}
    \begin{minipage}[t]{0.5\linewidth} 
        \centering
        \begin{adjustbox}{width=\linewidth} 
            \includegraphics[width=0.75\textwidth, left]{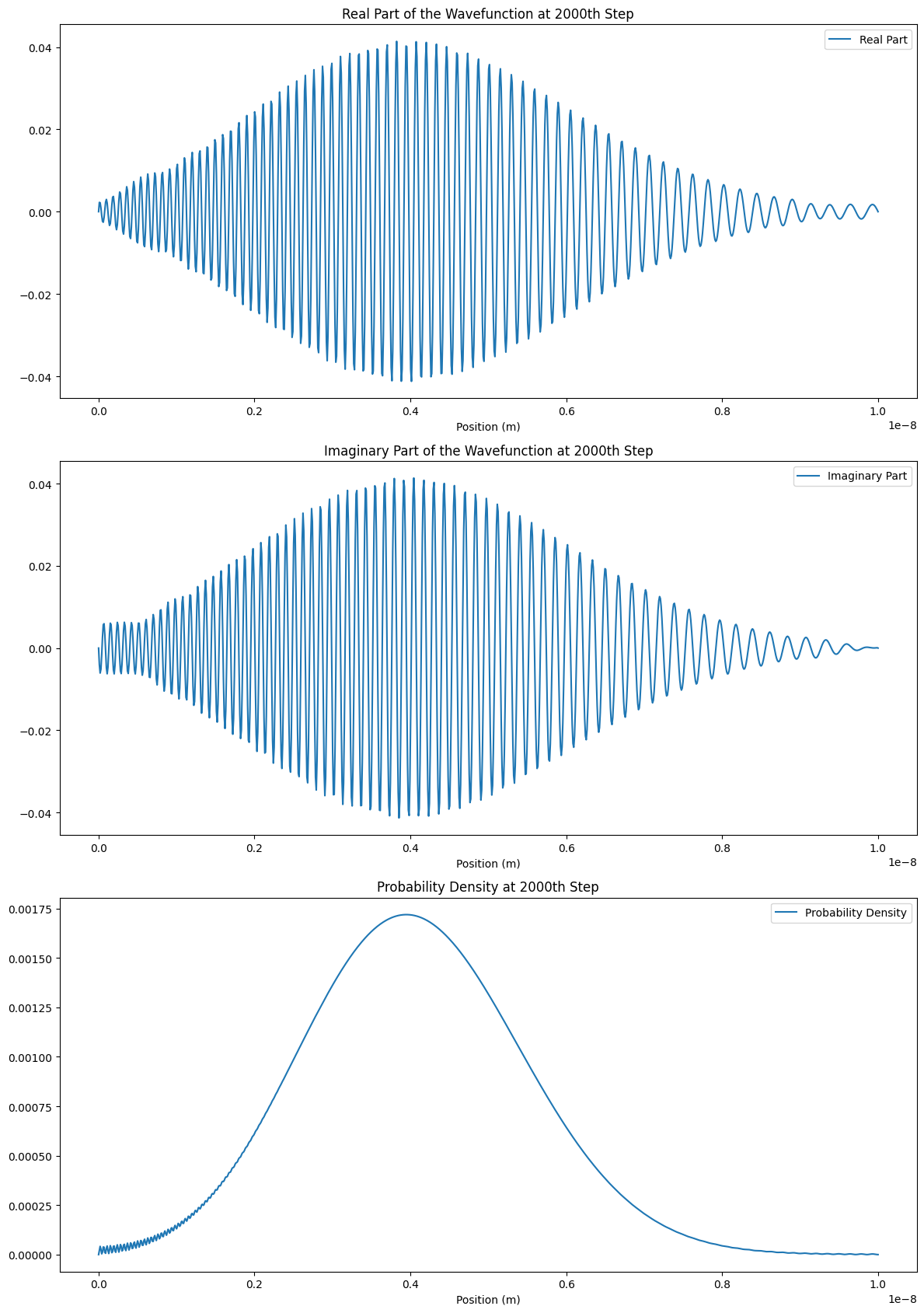} 
        \end{adjustbox}
    \end{minipage}
    \hfill
    \begin{minipage}[t]{0.5\linewidth} 
        \centering
        \begin{adjustbox}{width=\linewidth} 
            \includegraphics[width=0.8\textwidth, right]{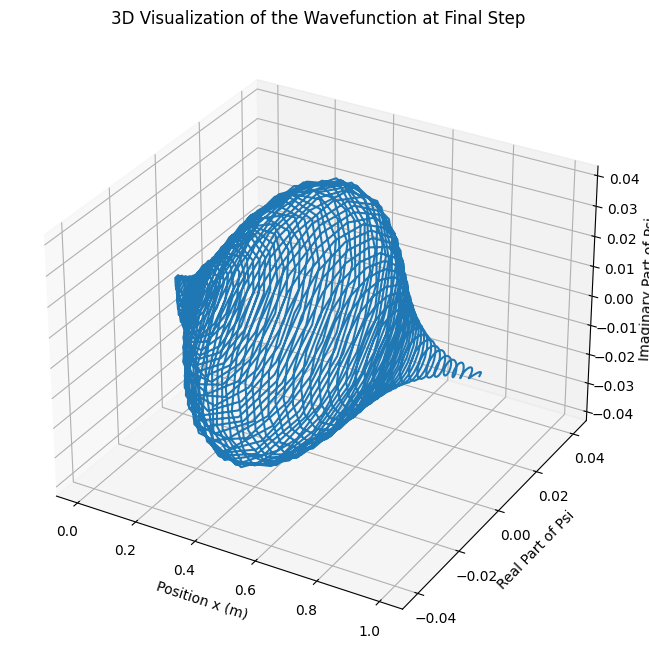} 
        \end{adjustbox}
    \end{minipage}
    \caption{The left diagram denotes the 2D-plots of real part, imaginary part, and probability density of the wavefunction with distance at the final time step. The right diagram shows the 3D evolution of the real and imaginary part of the wavefunction at the final time step}
    \label{fig:final-time-step}
\end{figure}


\begin{figure}
    \begin{minipage}[t]{0.5\linewidth} 
        \centering
        \begin{adjustbox}{width=\linewidth} 
            \includegraphics[width=0.75\textwidth, left]{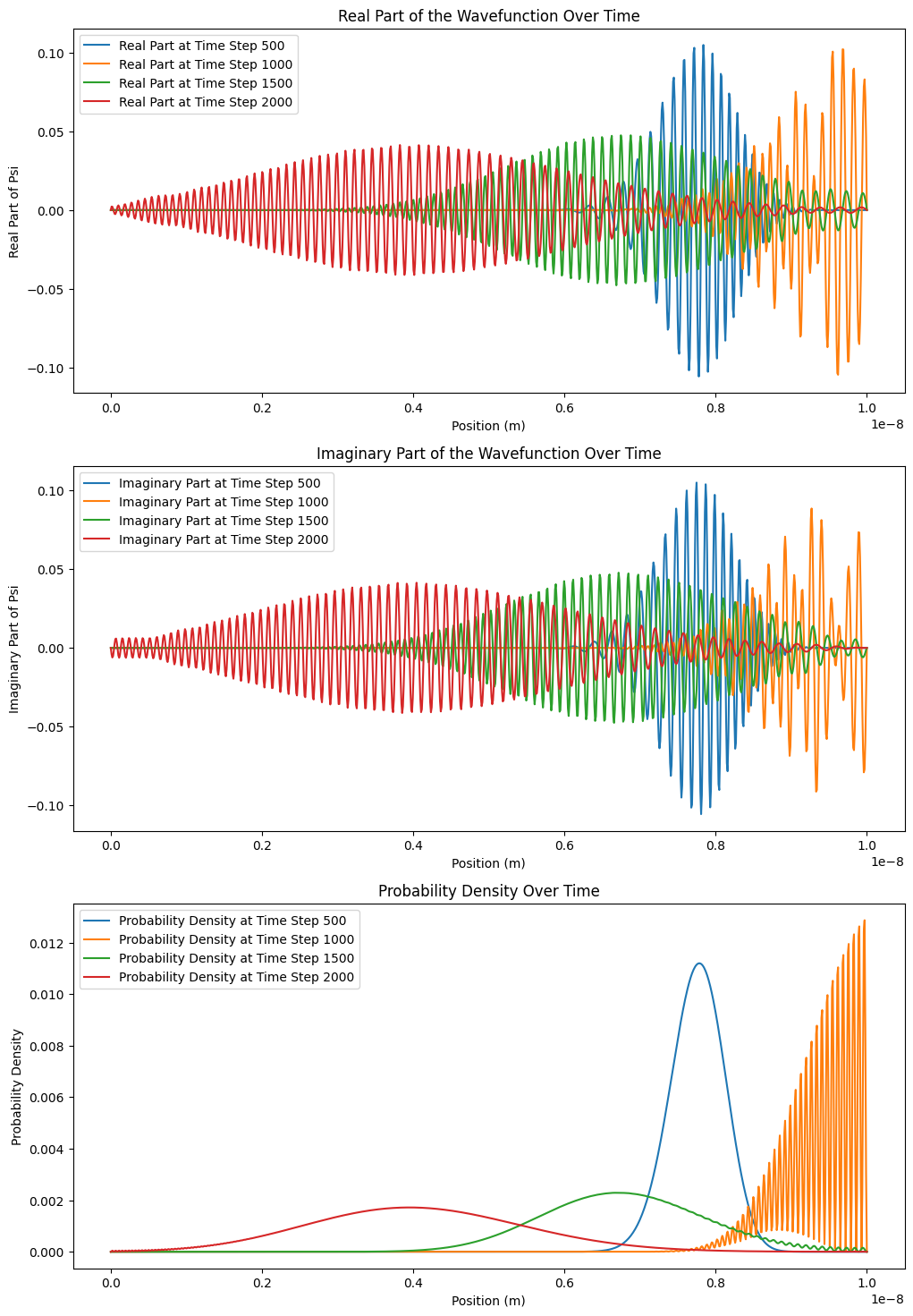} 
        \end{adjustbox}
    \end{minipage}
    \hfill
    \begin{minipage}[t]{0.55\linewidth} 
        \centering
        \begin{adjustbox}{width=\linewidth} 
            \includegraphics[width=0.8\textwidth, right]{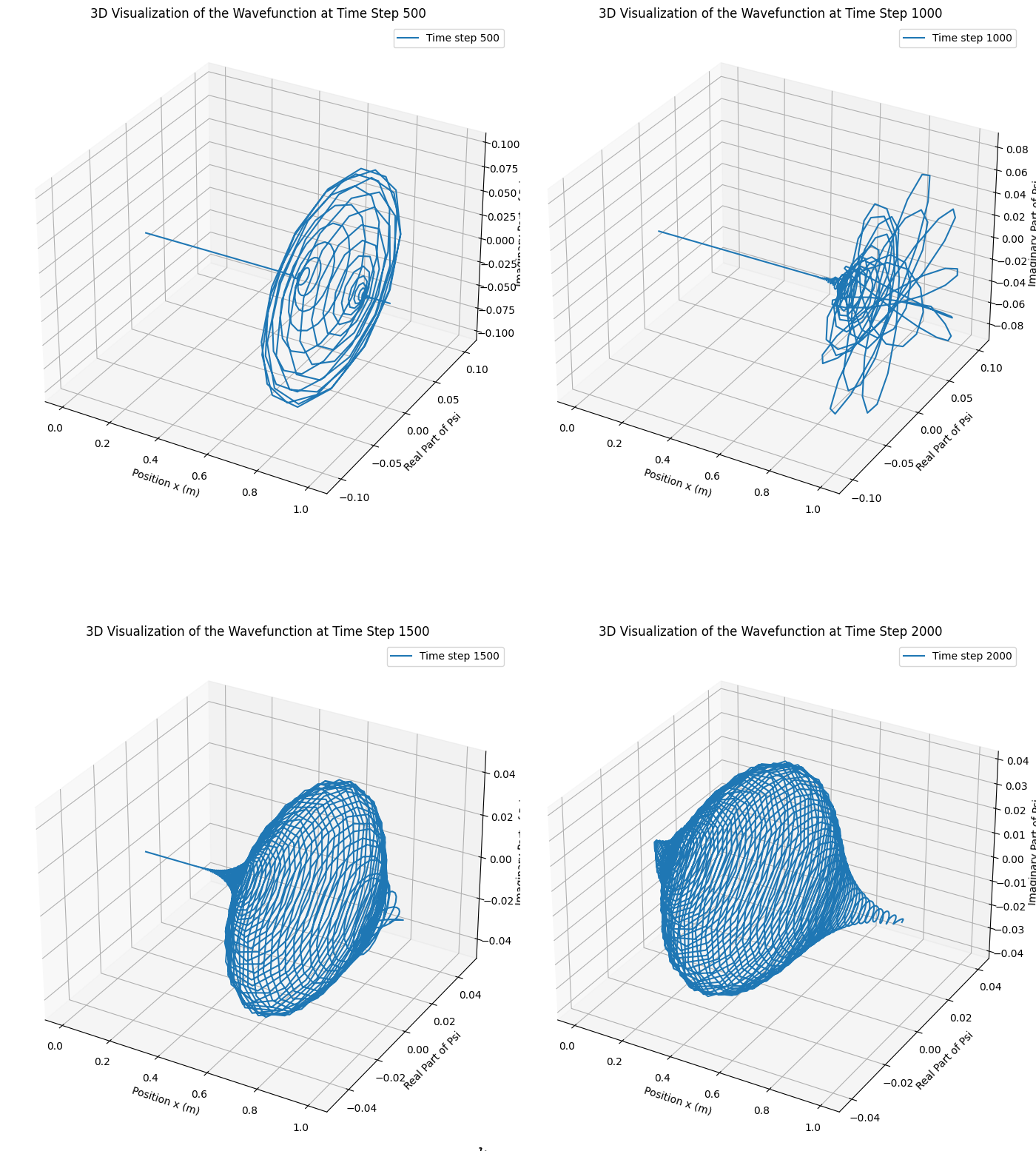} 
        \end{adjustbox}
    \end{minipage}
    \caption{The left diagram denotes the 2D plots of the real part, imaginary part, and probability density of the wavefunction with distance for a set of multiple time steps. The right diagram shows the 3D evolution of the real and imaginary part of the wavefunction for some multiple time steps}
    \label{fig:multiple-time-steps}
\end{figure}

\paragraph{}
The evolution of the real and imaginary parts of the wavefunction over time vividly illustrates the principle of quantum superposition and interference. From left diagram of figure 2, we see that at early time steps, such as 500 and 1000, the wave patterns are more coherent and less dispersed, indicating the wavefunction's initial response to the boundary conditions of the well. As time progresses to steps 1500 and 2000, the waveforms show increased complexity and spatial variation, reflecting the wavefunction's adaptation to the confining potential. This complexity is due to the superposition of multiple quantum states (eigenstates of the well) that contribute to the time-dependent wavefunction. The real part of the wavefunction appears as oscillatory patterns that shift over time, reflecting the particle’s momentum components. The imaginary part complements this by representing the phase of the wavefunction, crucial for understanding the full quantum state since the probabilities are derived from the complete wavefunction (combining real and imaginary parts).
\paragraph{}
The probability density plots are particularly insightful, demonstrating how the likelihood of finding the electron in particular regions of the well changes over time. Initially, the electron's presence is more likely in the middle of the well, as expected from a Gaussian initial state centered there. Over time, especially noticeable at the final step (2000), the probability density spreads out and shows peaks at specific intervals. This distribution is characteristic of the standing wave patterns formed by the wavefunction in a confined space, dictated by the quantum mechanical boundary conditions that require the wavefunction to be zero at the well's edges.
\paragraph{}
The 3D time visualization for four different time steps most convincing perspective of the wavefunction's evolution. Each snapshot at different time steps (500, 1000, 1500, and 2000) from 3D time evolution animation captures a three-dimensional plot combining both real and imaginary components. This visualization emphasizes the wave-particle duality nature of quantum mechanics, where the electron's behavior is not just a particle localized in space but rather a wave spread across the well. The evolution from one-time step to another showcases the wavefunction's oscillatory nature, where peaks and troughs evolve according to the superposition principle and the boundary-imposed constraints.

\section{Conclusion}

The project has successfully demonstrated the effectiveness of the Crank-Nicolson method for simulating the time evolution of a quantum particle's wavefunction within a confined potential well. The simulation results reproduced the expected quantum mechanical phenomena, such as wave interference and the formation of standing wave patterns within the potential well, adhering closely to theoretical predictions. Through proper implementation of this technique and detailed visualization of both the real, imaginary, and probability densities of the wavefunction, we have been able to observe and analyze quantum mechanical behaviors in a numerically stable manner. This project has, therefore, not only validated a computational approach but also enhanced the understanding of quantum mechanics in bounded systems. While the simulation accurately reflected the quantum phenomena of a single electron, there's still some room to introduce further improvement in this simulation. Future updates could explore different potential profiles, extend to multiple particles, and enhance performance for broader applications. Integrating the simulation with quantum computing platforms could also open up new possibilities for advanced quantum simulations. Overall, the project offers valuable insights and a solid foundation for further exploration in quantum mechanics simulations.
\paragraph{}
\paragraph{}
\paragraph{}
\section{References}

\begin{enumerate}
    \item Kabir, Adib. \url{https://colab.research.google.com/drive/16k0Z2nSO9ylePPLS81kjmGOuE7ZoFgjy#scrollTo=b_qgg0PFP_rm}. 
    \item Newman, Mark E. J. \textit{Computational Physics}.
    \item Griffiths, David J. \textit{Introduction to Quantum Mechanics}. 2nd ed., Pearson Prentice Hall, 2005.
    \item Kantor, Jack. \textit{Animation in Jupyter Notebooks}. GitHub, \url{https://colab.research.google.com/github/jckantor/CBE30338/blob/master/docs/A.03-Animation-in-Jupyter-Notebooks.ipynb}. Accessed 23 April 2024.
\end{enumerate}

\end{document}